\documentclass[twocolumn]{article}
\usepackage[utf8]{inputenc}
\usepackage{graphics}
\usepackage{authblk}
\usepackage{textgreek}
\usepackage{mathtools}
\usepackage{cleveref}
\usepackage{graphicx}

\begin{document}

\title{Interpretation of Dynamic Nanoindentation Results a simple harmonic oscillators for measurement of viscoelasticity}

\author[1]{Ed Darnbrough}

\affil[1]{%
Department of Materials, University of Oxford, 
Parks Road, Oxford OX1 3PH, UK. 
ed.darnbrough@materials.ox.ac.uk}

\twocolumn[
  \begin{@twocolumnfalse}
    \maketitle

%\begin{justify}

\section*{Abstract}

With the invent of nanoindentation technology capable of greater frequency of oscillation the full resonant behaviour can be observed. Here we lay out a proposed mathematical basis to interpret the measured dynamic compliance of a system to discover the viscoelastic properties of test samples. 

%\end{justify}

\vspace{2em}

\end{@twocolumnfalse}
]

\section*{Introduction}
The role of test frequency in the measured response and how to model the interaction between the oscillating indenter and the sample has been discussed in many papers \cite{Nix1998, Fischer-Cripps2004, Pharr2009, Leitner2017, Phani2020} with renewed interest in light of strain rate testing \cite{Soer2005,Li2011,Muthupandi2017} and the application of nanoindentation to time dependent displacement materials eg polymers and biomaterials \cite{Huang2004,Herbert2009, Hardiman2016,Wang2017b,Gao2017}. 

\section*{Test Method}
Tests are conducted with a berkovich shaped tip with a Hysitron PI88 low load system in a vacuum (1e-5 Pa) to remove any dampening effects that ambient air may introduce. An automated approach and drift test is conducted at low loads (20 microN) before loading the sample to a chosen amount where it is held in load control. During this load hold the frequency of oscillation is varied between 1 and 300 Hz in a non-sequential method to avoid any bias due to increasing depth throughout the test. Data is recorded for each frequency with a minimum number of oscillations. The results are then considered for each frequency taking the mean value and the standard deviation which are then represented as errorbars in the following plots. 

\section*{Mathematical Model}
We consider the indenter tip to be acting as a driven damped harmonic oscillator in series with further damped harmonic oscillators \cite{MechanicsTheory}. As the displacement of each consecutive oscillator is driven by the previous the magnitude decays rapidly and here we will only consider three in the figure below, \ref{fig:setup}, and our data analysis however the general formalization holds. 

Consider the sinusoidally driven position to be $x_{0}$ which can be represented as below, assuming at $t$=0 $x_{0}=0$:
\begin{center}
$k_{1}x_{0} + b_{1}x_{0}' = F e^{i \omega t}$ \\
$x_{0} =\frac{F e^{i \omega t}}{k_{1}+ib_{1}w} $ \\
\end{center} 

\begin{figure}[h]
    \centering \includegraphics[width = 0.25\textwidth]{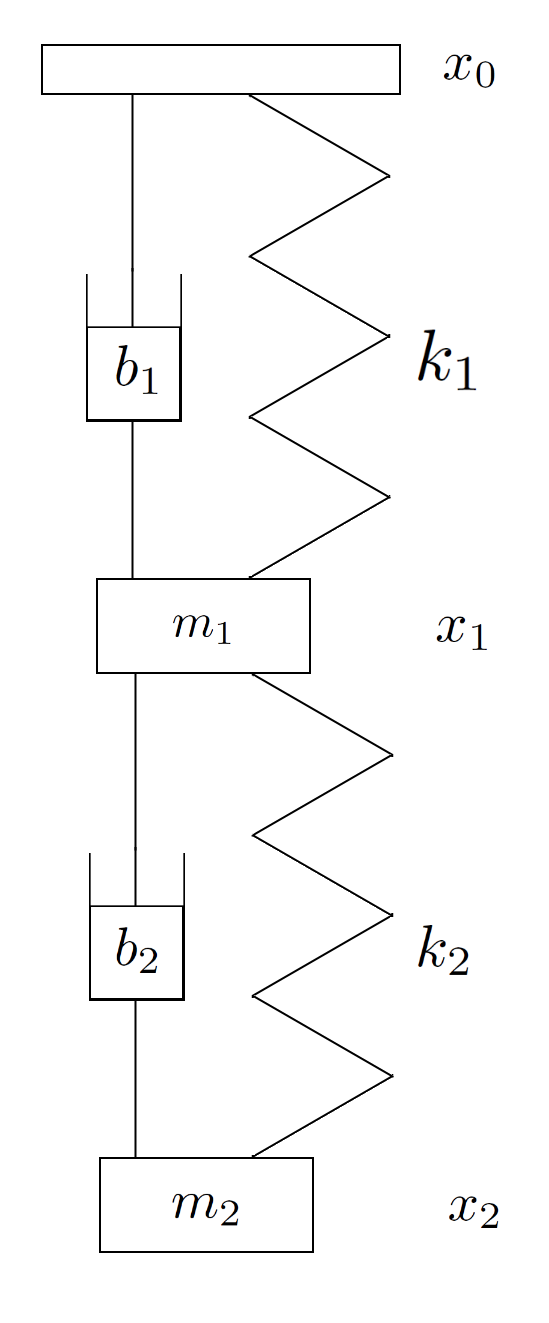}
    \caption{A schematic of an example damped spring system where $m_{1}$ can be considered the effective mass of the tip and $m_{2}$ the effective mass of the sampled material. }
    \label{fig:setup}
\end{figure}

then $x_{0}$ and the displacement of the first mass $x_{1}$ (in our case the tip) to be described by the equation: \\
$b_{1}(x_{0}'-x_{1}') + k_{1}(x_{0}-x_{1}) = m_{1}x_{1}'' + b_{2}x_{1}' + k_{2}x_{1} $ \\
This can be seperated in to terms relating to the driver and driven: \\
$m_{1}x_{1}''+ (b_{1}+b_{2})x_{1}' + (k_{1}+k_{2})x_{1} = k_{1}x_{0} + b_{1}x_{0}'$ \\

If we assume that $x_{1}$ will also be of a sinusoidal form then the equation can be complexified and a characteristic polynomial used to find a result of the form:
\begin{center}
$p_{1}(r)\tilde{x_{1}} = (k_{1}+b_{1}i \omega)\tilde{x_{0}}$ \\
$\tilde{x_{1}} = \frac{k_{1}+ b_{1}i \omega}{p_{1}(i\omega)}\tilde{x_{0}} $\\
where \\
$p_{1}(i\omega) = (k_{1}+k_{2}) + i\omega(b_{1}+b_{2}) - \omega^{2}m_{1}$ 
\end{center}
This means that the amplitude of the sinusoidal motion of $\tilde{x_{1}}$ is related to $\tilde{x_{0}}$ by a factor dependent on the frequency, stiffness of both springs, mass and damping coefficients called gain, $\tilde{g}$. Therefore it can be seen that the motion of subsequent masses can be written in a general form: 
\begin{center}
$\tilde{x_{j}} = \prod_{1}^{n} \tilde{g_{j}}\tilde{x_{0}} $ \\
$\tilde{g_{total}} = \sum_{1}^{n} \tilde{g_{j}} $
\end{center}
The gain for each mass depends on the components neighbouring it so a simple case with damped springs either side is: 
\begin{center}
$\tilde{g_{j}} = (k_{j}+i\omega b_{j})/p_{j}(i\omega)$ \\
$p_{j}(i\omega) = (k_{j}+k_{j+1}) + i\omega(b_{j}+b_{j+1}) - \omega^{2}m_{j}$
\end{center}
The second alternative we consider in this work is the mass bounded by a damped spring on one side and a dashpot on the other. 
\begin{center}
$\tilde{g_{j}} = (i\omega b_{j})/p_{j}(i\omega) $ \\
$p_{j}(i\omega) = (k_{j}) + i\omega(b_{j}+b_{j+1}) - \omega^{2}m_{j}$ 
\end{center}
To consider this set up with alternatives such as a Maxwell material is trivial and so is left as an exercise for the reader.  

To make use of this model we consider the complex gain to be made of a phase component and a magnitude: 
\begin{center}
$\tilde{g_{j}} = |\tilde{g_{j}}| e^{-i\phi}$ \\
$|\tilde{g_{j}}| = \frac{\sqrt{k_{j}^2+(\omega b_{j})^2}}{\sqrt{(k_{j}+k_{j+1} -m_{j}\omega^2)^2 + (\omega(b_{j}+b_{j+1}))^2}}  $ \\
$\phi = -Arg(\tilde{g}) = -atan2(Im, Real) = -arctan(\frac{\omega(b_{j}+b_{j+1})}{(k_{j}+k_{j+1} -m_{j}\omega^2)})$
\end{center}
It should be noted that the number of resonant frequencies available is determined by the number of effective masses within the total system \cite{}. A single mass will give only a single resonant peak but above that the mathematics becomes more complex and requires the equation of motion of each mass and solving the determinant for $\omega_{r}$. An example of this for the first two masses considered in this work without the driver so taking $x_{0} = 0$:    
\begin{center}
$m_{1}x_{1}''+ b_{1}x_{1}' + k_{1}x_{1} + b_{2}(x_{1}'-x_{2}') + k_{2}(x_{1}-x_{2}) = 0 $ \\
$m_{2}x_{2}''+ b_{2}(x_{2}'-x_{1}') + k_{2}(x_{2}-x_{1}) = 0$ \\

$\begin{vmatrix}
-b_{2}-k_{2} & k_{1}+k_{2} + \omega(b_{1}+b_{2}) + \omega^{2}m_{1} \\
k_{2} + \omega(b_{2}) + \omega^{2}m_{2} & -k_{2}-b_{2}
\end{vmatrix} = 0$

\end{center}
This method of solving linear second order differential equations will find all roots however given our experiments are limited to real and positive values of $\omega$ we can use the Descartes' Sign Rule \cite{}. One sign change per equation suggests there are two positive resonances in the system, or one for each mass. 

If we consider the subsequent mass and spring systems to have a summative effect on the measured dynamic compliance the effect of angle shows that at a frequency below resonance of a system $\phi$ is small so the contribution will be positive but past resonance $\phi$ is close to $\pi$ meaning it is almost completely out of phase and so a negative contribution. At resonance the angle will be $\pi/2$ making the sinusoidal oscillation a cosine form. This is interesting when considering the response of two systems (x and y) at the resonant frequency of one (x), $\omega_{rx}$, but before the resonant frequency of the other (y):
\begin{center}
$\Sigma g = g_{x}cos(\omega_{rx}) + g_{y}sin(\omega_{rx})$ \\
		$= \sqrt{g_{x}^2+g_{y}^2}(\frac{g_{x}}{\sqrt{g_{x}^2+g_{y}^2}}cos(\omega_{rx})+\frac{g_{y}}{\sqrt{g_{x}^2+g_{y}^2}}sin(\omega_{rx})) $\\
using: $sin(\alpha) = \frac{g_{x}}{\sqrt{g_{x}^2+g_{y}^2}} $ and $cos(\alpha) = \frac{g_{y}}{\sqrt{g_{x}^2+g_{y}^2}}$ \\
$= \sqrt{g_{x}^2+g_{y}^2}sin(\omega_{rx} + arctan(\frac{g_{x}}{g_{y}}))$
\end{center}
This means that the measured phase lag of the total system as measured at resonance is related to $arctan(\frac{g_{x}}{g_{y}})$ and the amplitude of gain is $\sqrt{g_{x}^2+g_{y}^2}$. This can be used with other known conditions, shown below, to make a first selection of variables when fitting real data. 
\begin{center}
at $\omega=0 \qquad  |g|= g_{x}+g_{y} = g_{1}(1+g_{2}) = \frac{k_{1}}{\sqrt{k_{1}^2+k_{2}^2}}(1+ k_{1})$\\
at $\omega=\omega_{r1} \qquad |g_{1}|= \frac{\sqrt{k_{1}^2+(\omega b_{1})^2}}{\sqrt{b_{1}^2+b_{2}^2}}$ \\
at $\omega=\omega_{r2} \qquad |g_{2}|= \frac{\sqrt{k_{2}^2+(\omega b_{2})^2}}{b_{2}}$
\end{center}

These points in order along with the understanding that the resonant frequency of a system is $\omega_{r1}^2 = \frac{k_{1}+k_{2}}{m_{1}}$ gives easy first guess for suitable values of stiffness, damping coefficient and mass. This is useful given the interconnection of the three variables as seen that increasing all of them by the same factor results in no change in the $|g(\omega)|$.

The stiffness and damping coefficient can be converted to global elastic and viscoelastic material values through the projected area of indentation. 
\begin{center}
$ E = \frac{\sqrt{\pi}}{2}\frac{k}{\sqrt{A(h)}}   \qquad  \nu = \frac{\sqrt{\pi}}{2}\frac{b}{\sqrt{A(h)}} $  \\
$ \tau_{r} = \frac{\nu}{E} = \frac{b}{k} $ 
\end{center}
Where the retardation time $\tau_{r}$ is independent of depth. 

\section*{Results}

\emph{Nota Bene: For ArXiv this section will be updated as data from different materials becomes available.}

The data collected using the method outlined above is then fitted using a Matlab script which extracts the values at $\omega=0$ and resonance to make an initial guess at the materials stiffness, damping coefficient and effective mass. From this guess a fitting routine based on the least squared residuals method created by Prof Des McMorrow and collaborators at I.L.L. is used allowing all variables to be optimised. The scripts used and the full data sets can be obtained through correspondence with the author.

\begin{figure}[h]
    \centering \includegraphics[width = 0.45\textwidth]{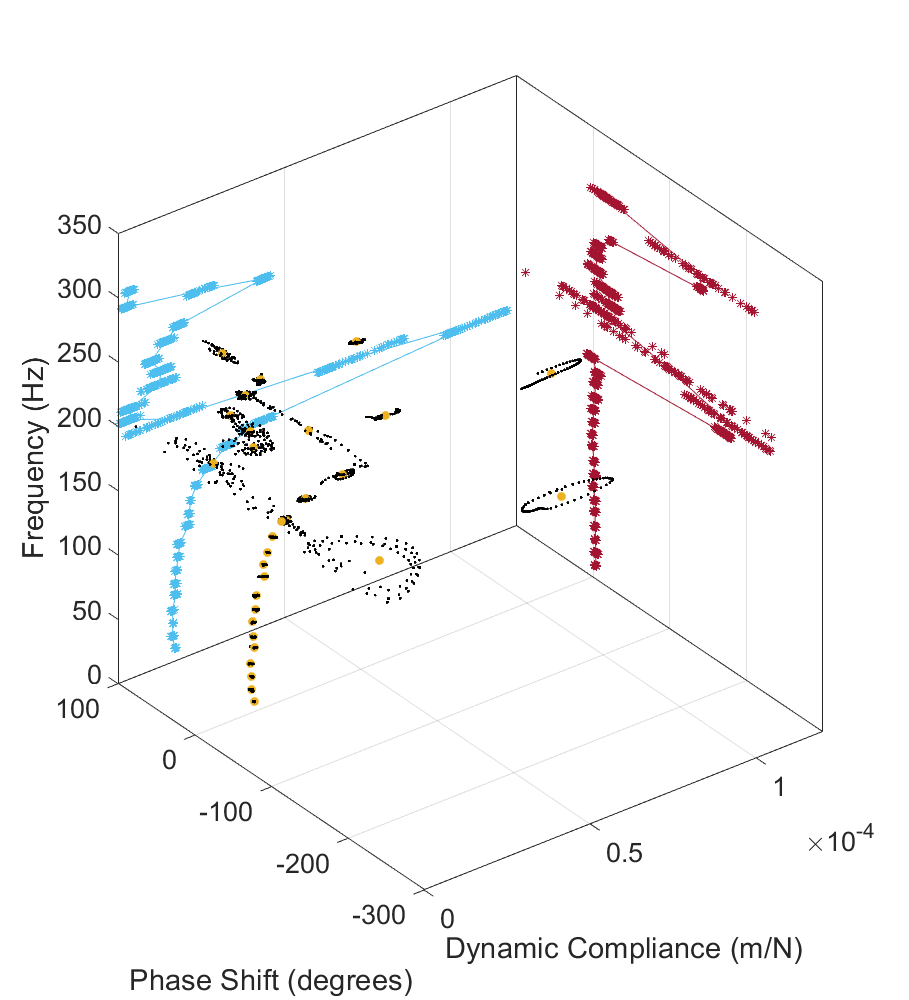}
    \caption{A typical data set from aluminium. }
    \label{fig:result}
\end{figure}

\begin{figure}[h]
    \centering \includegraphics[width = 0.455\textwidth]{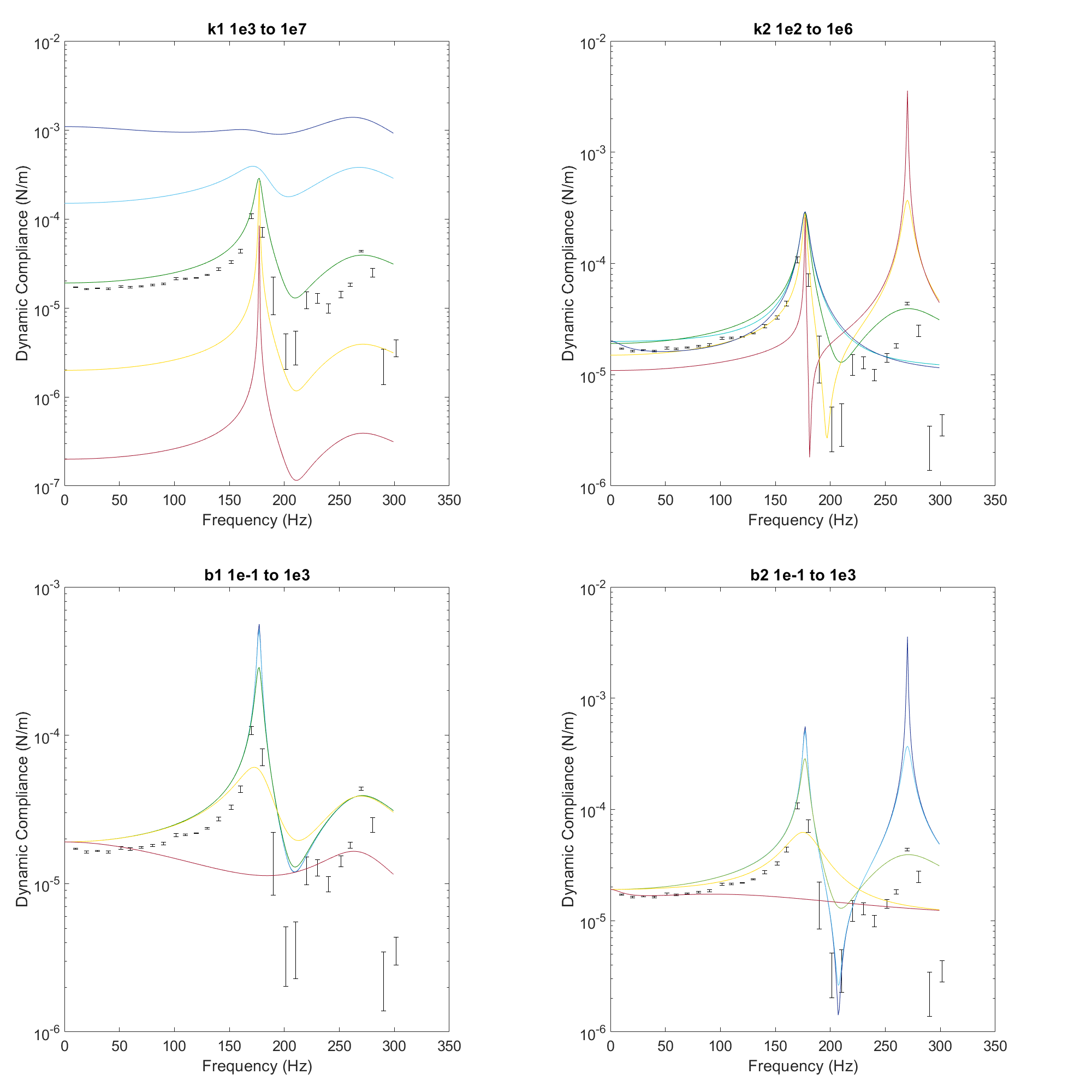}
    \caption{A sensitivity analysis illustrating the effect of different fitted parameters. }
    \label{fig:sensitive}
\end{figure}

The figure \ref{fig:result} shows a typical data set (in black) with the mean value of each frequency (in yellow) with 2D representations of the phase (in red) and the dynamic compliance (in blue). The fitting of such data can be done in either a complex space or considering just a single dimension. Choosing to fit predominately to the dynamic compliance given the reduced experimental scatter in that dimension, the routine described above is given a scripted guess and then the final fitting by routine with the same variables for the fitting in both the dynamic compliance and the phase plots. This illustrates that the model describes the system well as it encompasses features in both real space and phase space. The elastic values for the materials match literature values by other methods and then this allows confidence in the damping coefficient which in turn gives understanding of the viscoelastic response in the retardation time.
To give the reader appreciation of the role of each of the fitted parameter figure \ref{fig:sensitive} shows a sensitivity analysis with typical data (in black) a human given guess (in green) and then each of the four key variables changed by two orders of magnitude up (yellow and red) and down (blue and dark blue) in turn to give the resultant plots.

\section*{Discussion}
One question to contemplate is the role, and physical meaning, of effective mass during an experiment. The size of the effective mass can effect the breadth of the resonance observed and the response can differ between different test samples. This could be understood as different volumes of material being interacted with dependent on the size of the stress field. The observed changes to the tip effective mass is possible that the amount of surface traction between the tip and the material effects the apparent weight due to an additional force.

\section*{Conclusions}
This paper lays out an alternate and complementary description of dynamic nanoindentation experimental system to extract a material's viscoelastic properties independently of machine compliance. The above detailed two series component system adequately describes the results seen experimentally with the theory suggesting that additional components would have reducing effect there is no further need to consider them.   

\section*{Acknowledgements}
This work is a complementary offshoot of research conducted as part of the SOLBAT project funded by the UK Faraday Institute on equipment funded by the Royce Institute. 

\bibliographystyle{unsrt}
%\bibliography{StrainrateElastic}

\end{document}